\documentclass[twocolumn]{aastex62}

\usepackage{graphicx}	
\usepackage{amsmath}	
\usepackage{amssymb}	
\usepackage{bm}		
\usepackage{multirow}

\usepackage{placeins}

\received{8 May 2019}
\revised{27 May 2019}
\accepted{xx}
\submitjournal{ApJL}

\shorttitle{Disk-jet coupling in MAXI~J1535$-$571 during re-brightenings}
\shortauthors{Parikh {\em et al.}}

\newcommand{\mx}{MAXI J1535$-$571}
\newcommand{\nh}{$N_\mathrm{H}$}
\usepackage{amssymb}
\usepackage{pifont}
\usepackage{xcolor}
\usepackage{seqsplit}

\begin{document}

\title{Rapidly evolving disk-jet coupling during re-brightenings in the black hole transient MAXI~J1535$-$571}

\author{A. S. Parikh}
\affiliation{Anton Pannekoek Institute for Astronomy, University of Amsterdam, Postbus 94249, NL-1090 GE Amsterdam, the Netherlands}

\author{T. D. Russell}
\affiliation{Anton Pannekoek Institute for Astronomy, University of Amsterdam, Postbus 94249, NL-1090 GE Amsterdam, the Netherlands}

\author{R. Wijnands}
\affiliation{Anton Pannekoek Institute for Astronomy, University of Amsterdam, Postbus 94249, NL-1090 GE Amsterdam, the Netherlands}

\author{J. C. A. Miller-Jones}
\affiliation{International Centre for Radio Astronomy Research, Curtin University, GPO Box U1987, Perth, WA 6845, Australia}

\author{G. R. Sivakoff}
\affiliation{Department of Physics, University of Alberta, Edmonton, AB T6G 2G7, Canada}

\author{A. J. Tetarenko}
\affiliation{Department of Physics, University of Alberta, Edmonton, AB T6G 2G7, Canada}
\affiliation{East Asian Observatory, 660 N. A`ohoku Place, University Park, Hilo, HI 96720, USA}


\email{a.s.parikh@uva.nl}


\begin{abstract}
The main outburst of the candidate black hole low-mass X-ray binary (BH LMXB) \mx\, ended in 2018 May and was followed by at least five episodes of re-brightenings. We have monitored this re-brightening phenomenon at X-ray and radio wavelengths using the {\it Neil Gehrels Swift Observatory} and Australia Telescope Compact Array, respectively. The first two re-brightenings exhibited a high peak X-ray luminosity (implying a high mass accretion rate) and were observed to transition from the hard to the soft state. However, unlike the main outburst, these re-brightenings did not exhibit clear hysteresis. During the re-brightenings, when \mx\, was in the hard state, we observed the brightening of a compact radio jet which was subsequently quenched when the source transitioned to a similar soft state as was observed during the main outburst. We report on the first investigation of disk-jet coupling over multiple rapidly evolving re-brightenings in a BH LMXB. We find that the accretion flow properties and the accompanying compact jet evolve on a similarly rapid time scale of $\sim$days rather than the typical value of $\sim$weeks as observed for most other BH LMXBs during their main outburst events. 
\end{abstract}


\section{Introduction}
\label{introduction}

Low-mass stars in close binaries, with black holes (BHs) and neutron stars, can overflow their Roche lobes, and are commonly referred to as low-mass X-ray binaries (LMXBs). The overflowing companion material forms a disk around the compact object. Instabilities in this disk can increase the mass accretion rate ($\dot{M}$) onto the compact object, causing `outbursts' \citep[e.g.,][]{osaki1974accretion,lasota2001disc}. 

Accretion outbursts in BH LMXBs typically span orders of magnitude in luminosity during which the source may broadly exhibit two main spectral states -- hard and soft. During the outbursts, the observed X-ray luminosity serves as a proxy for the $\dot{M}$. As the outburst begins, the luminosity rises (indicating an $\dot{M}$ increase) and the source is initially in a power-law dominated X-ray `hard' state (HS). 
As the source continues to brighten further (with increasing $\dot{M}$) the source transitions to a disk-dominated X-ray `soft' state (SS). 
Once the outburst begins to decay, as $\dot{M}$ decreases, the luminosity drops and the source returns to the HS. Observations indicate that the luminosity at which the source transitions from the soft-to-hard state during the outburst decay tends to be lower than the luminosity of the hard-to-soft state transition during the outburst rise, exhibiting a hysteresis \citep[e.g.,][]{miyamoto1995large,belloni2005evolution,meyer2005hysteresis,dunn2010global,vahdat2019investigating}. 

The HS in BH LMXBs is accompanied by a highly collimated outflow in the form of a compact jet. The observed radio emission traces the evolution of the jet. This partially self-absorbed synchrotron emitting jet exhibits a flat or inverted spectrum in the radio through infrared \citep[corresponding to a spectral index $\alpha \!\gtrsim\!0$; $S_{\nu}\!\propto\! \nu^{\alpha}$ where $\nu$ is the observing frequency and $S_{\nu}$ is the observed flux density;][]{fender2001powerful}. The jet has been observed to be quenched in some systems by $\gtrsim$2.5 orders of magnitude as the source transitions to the SS \citep[e.g.,][]{fender1999quenching,russell2011testing}. The physics behind the jet launching and quenching process, as well as the connection to the accretion flow are not well understood. 

BH LMXBs have been found to exhibit a non-linear relationship between the radio (at 5 GHz) and X-ray luminosities (for the 1--10 keV energy range; $L_\mathrm{R} \propto L_\mathrm{X}^{\beta}$), indicating that the inflow traced by the X-ray luminosity and outflow traced by the radio luminosity are correlated \cite[][]{gallo2012assessing,corbel2013universal}. This relationship is defined during the HS data and is manifested as two distinct `radio-loud' and `radio-quiet' tracks for which $\beta  \! \sim  \! 0.6$ and $\beta  \! \sim  \! 1.0$, respectively \citep{gallo2012assessing}. It is not certain why BH LMXBs trace out these two tracks. Some sources are found to trace both tracks, moving from one to another as the luminosity evolves above or below a certain value during their outburst \citep[e.g., H1743$-$322, MAXI~J1659$-$152, Swift~J1753.5$-$0127;][]{coriat2011radiatively,jonker2012black,plotkin2017up}. More recent work taking into account a larger sample of BH sources, however, suggests that the two tracks described have not proven to be robustly partitioned \citep[e.g.,][]{gallo2014radio,gallo2018hard}.

In addition to the main accretion outburst, some compact binary systems exhibit `re-brightenings' after the end of their main outburst. They are seen in systems hosting BHs, neutron stars, and white dwarfs \citep[e.g.,][]{osaki2001repetitive,patruno2016reflares,yan2017detection}. The re-brightening behaviour across these systems appears very similar, which suggests that the mechanism may primarily be influenced by the accretion physics in the disk rather than the compact object itself. The re-brightening phenomenon cannot be explained within the disk instability model in a straightforward manner \citep{dubus2001disc,lasota2001disc} and its physical origin is not known. Evidence suggests that it may be triggered by the same hydrogen ionisation instability that initiates regular outbursts \citep{patruno2009saxj1808} and/or it may be triggered by increased irradiation by the donor \citep{hameury2000zoo}.

\mx\,is a candidate BH LMXB that was detected independently by the {\it Neil Gehrels Swift Observatory} (hereafter {\it Swift}) and the {\it Monitor of All-sky X-ray Image (MAXI)} in outburst in early 2017 September \citep{barthelmy2017GCN,negoro2017maxi}. Its main outburst ended in 2018 May, after which the source exhibited a series of re-brightening events, seen at radio and X-ray frequencies \citep[e.g.,][]{parikh2018rebrightening}. The phenomenological behaviour of these re-brightenings is not consistent with any of the classifications reported in \citet{zhang2019bright}. The radio properties of BH LMXBs during such re-brightenings are very poorly studied. In this letter, we report on our \seqsplit{quasi-simultaneous} X-ray and radio coverage of the multiple re-brightenings in \mx\, and investigate its disk-jet coupling behaviour. It is the first time that the simultaneous coverage of such rapidly evolving re-brightenings has been studied with a high observing cadence.\footnote{We note that another BH LMXB Swift J1753.5$-$0127 was also studied in the X-ray and radio during its mini-outbursts by \cite{plotkin2017up}. However, these mini-outbursts occurred after a $\sim$3 month quiescent period following the main outburst. Thus, studying the re-brightenings in our source examines a different phenomenology.}

\section{Observations, Data Analysis, and Results}
\mx\,was observed using {\it Swift} in the X-ray regime and using the Australia Telescope Compact Array (ATCA; project code C3057) in the radio regime during both its main outburst and its re-brightenings. Here we report on the data obtained during the re-brightenings and we refer to \citet{russell2019maxi} for the results obtained for the main outburst. 
All errors reported here correspond to the 1$\sigma$ confidence range.

\subsection{\textit{Swift}}

\begin{figure}[h!]
\centering
\includegraphics[scale=0.46]{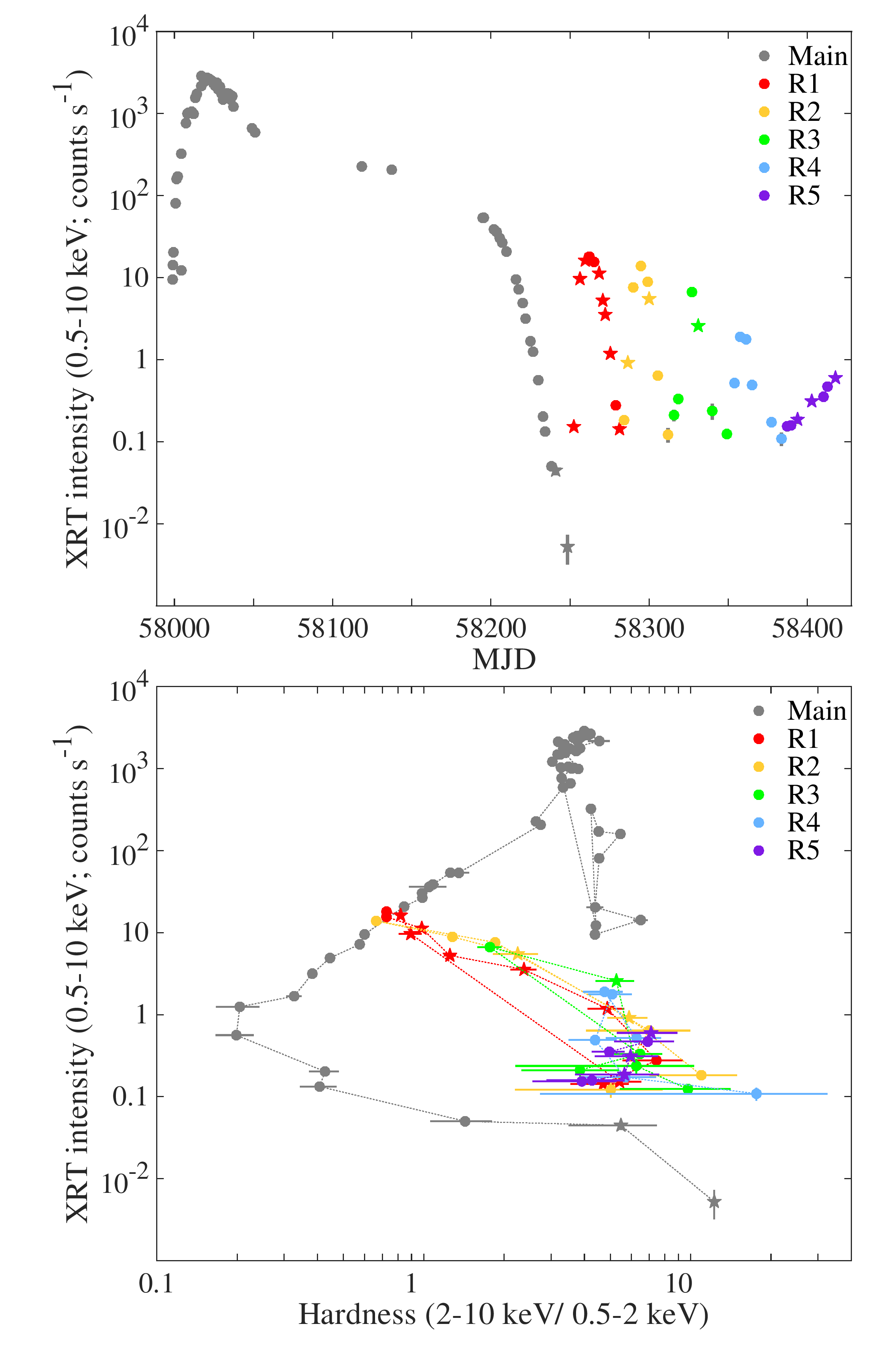}
\caption{The top panel shows the (XRT background-subtracted, 0.5--10 keV) light curve of \mx\,during its main outburst (shown in grey) and the subsequent re-brightenings (each indicated by a different colour). The bottom panel shows the hardness-intensity diagram for the main outburst and the re-brightenings (employing the same colour scheme as adapted for the light curve in the upper panel). The data points indicated using a star have quasi-simultaneous radio data, obtained using ATCA.}
\label{fig_lc_hid}
\end{figure}

\mx\,was observed 105 times during its main outburst and subsequent re-brightenings using {\it Swift} before the source became Sun constrained at the end of 2018. The last observation studied here was carried out on 2018 October 26. The data were obtained in both the Photon Counting (PC) and Windowed Timing (WT) modes of the X-ray Telescope (XRT). The raw data were downloaded from the \textsc{HEASARC} archive and were processed using the \textsc{xrtpipeline} tool. \textsc{XSelect} (v2.4d) was used to analyse the data and extract the count rates and spectra from the various observations. For this extraction, a circular region of 30 arcsec centred on the source was used.\footnote{We have adapted the \texttt{BACKSCAL} keyword (see http://www.swift.ac.uk/analysis/xrt/backscal.php) in the WT mode spectra, as appropriate, since we have used two dimensional extraction regions.} The background region used was annular, with an inner and outer radius of 200 and 300 arcsec, respectively. The observations affected by pile-up were corrected by discarding data corresponding to the piled-up central region of the source.\footnote{http://www.swift.ac.uk/analysis/xrt/pileup.php}

\begin{figure*}[ht]
\centering
\includegraphics[scale=0.46]{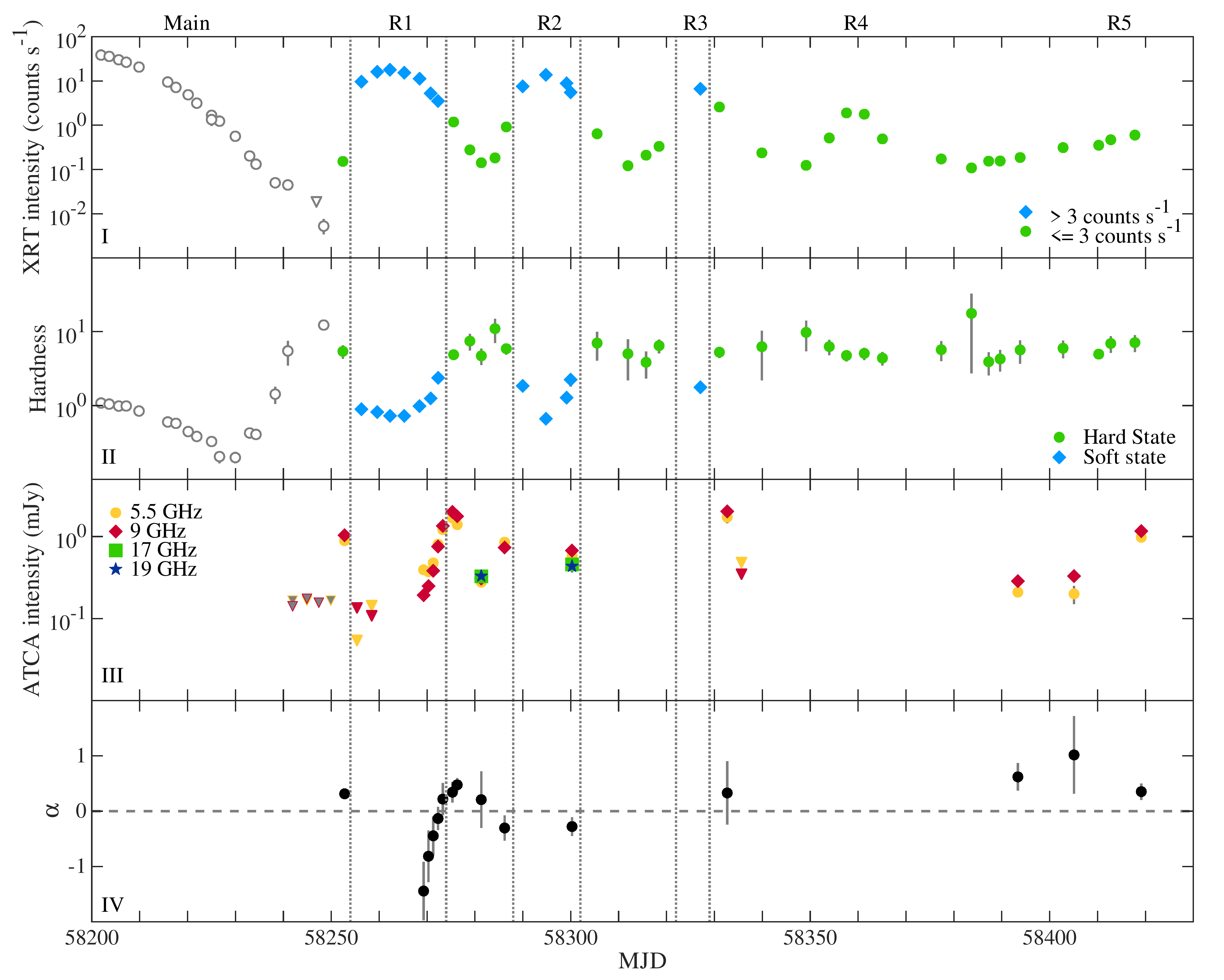}
\caption{Panel I shows the X-ray light curve evolution of \mx, observed using {\it Swift} (see also Figure \ref{fig_lc_hid}, upper panel). Representative points from the main outburst are shown in grey in Panels I and II of this figure. The vertical dotted grey lines roughly indicate the times of observed X-ray state transitions. Data corresponding to a source intensity $>$3 counts s$^{-1}$ and  $\leq$3 counts s$^{-1}$ are shown using blue $\textrm{\ding{117}}$ and green $\bullet$, respectively, in Panel I. Panel II shows the hardness evolution of the source with the blue $\textrm{\ding{117}}$ and green $\bullet$ indicating the soft and hard data (i.e., hardness $>$3 or $<$3), respectively. Panel III shows the radio fluxes in the 5.5, 9, 17, and 19 GHz bands (detections shown using yellow $\bullet$, red $\textrm{\ding{117}}$, green $\textrm{\ding{110}}$, and dark blue $\textrm{\ding{78}}$, respectively; upper limits are always shown using {\color{gray} $\textrm{\ding{116}}$} in the appropriate colour) obtained using ATCA. Panel IV shows the radio spectral index $\alpha$ determined using the ATCA data.}
\label{fig_mjd_r_x}
\end{figure*}

The background-subtracted light curve, extracted for the 0.5--10 keV energy range, is shown in the top panel of Figure \ref{fig_lc_hid}. The different colours are used to indicate the main outburst and the different re-brightenings. The light curve (using the same data as shown in the top panel of Figure \ref{fig_lc_hid}, but employing a different colour scheme) of the end of the main outburst and the subsequent re-brightenings is shown in Panel I of Figure \ref{fig_mjd_r_x}. The faintest luminosity level was achieved by the source after the end of the main outburst and before the first re-brightening, when it exhibited a count rate of $5.3 \times 10^{-3}$ counts s$^{-1}$. Once the source started re-brightening, the faintest luminosity level exhibited by \mx\,in between the different re-brightenings corresponded to $\sim$0.1--0.2 counts s$^{-1}$. The peak of the first re-brightening was a factor of $\sim$160 lower in brightness than the peak of the main outburst. Furthermore, the peak brightness of the re-brightenings decayed with time --- evolving from $\sim $18 counts s$^{-1}$ to $\sim$1.9 counts s$^{-1}$ over the initial four re-brightenings. The first four re-brightenings were observed across their entire evolution from rise to decay and exhibited a similar behaviour with respect to each other. They all displayed a steep sharp rise compared to the preceding decay trend, accompanied by a steep decay after the peak of the re-brightening was reached. The peak of each these four re-brightenings are separated by a period of 33$\pm$1 days. The peak of the last (fifth) re-brightening is not known as the source could no longer be observed due to Sun constraint. However, its rise was observed and was found to be much slower than that seen for the previous four re-brightenings, hinting at a different evolution.

Comparing the emission of the source in the soft (0.5--2 keV) and hard energy bands (2--10 keV) indicates the energy range in which the emission is dominant and thereby the spectral state exhibited by the source. We use the hardness ratio (the ratio of counts in the hard band to those in the soft band) to study this. The hardness--intensity diagram (HID) is shown in the lower panel of Figure \ref{fig_lc_hid}. Once again the colours in the HID correspond to the main outburst and the 5 subsequent re-brightenings studied here. The hardness evolution with time is shown in Panel II of Figure \ref{fig_mjd_r_x}.

Examining the HID shows that the main outburst transited to the SS and traced out a hysteresis loop \citep[see e.g.,][for an HID based on the {\it MAXI} data]{tao2018swift}. Only the first two re-brightenings transited to the SS (reaching similar low values of the hardness as seen during the main outburst), as can be seen in the HID. However, unlike the main outburst, these two re-brightenings did not appear to exhibit clear hysteresis. Interestingly, they were found to transit back to the HS following the same track in the HID as traced out during the transition to the SS. As the source intensity during the re-brightenings exceeds  $\sim$3 counts s$^{-1}$ the hardness ratio drops below $\sim$3 and the source transitions out of the HS (and vice versa for the complementary transition into the HS). This can also be seen in the top two panels of Figure \ref{fig_mjd_r_x}. The data corresponding to $>$3 counts s$^{-1}$ are shown using blue diamonds and those $\lesssim$3 counts s$^{-1}$ are shown using green circles. 
The HID of the third re-brightening indicates that it transited out of the HS but did not reach the hardness level corresponding to the SS determined from the main outburst and the first two re-brightenings. The fourth and fifth re-brightenings did not exceed $\sim$3 counts s$^{-1}$ and did not transit out of the HS.

\subsection{ATCA}

\mx\,was observed 22 times using ATCA during the re-brightenings (before the source was Sun-constrained) at the central frequencies of 5.5 and 9 GHz, and at an additional two frequencies of 17 and 19 GHz on 2018 June 12 and 2018 July 1. Details about the observation set-up and data reduction can be obtained from \citet{russell2019maxi}. The flux of the source in the radio bands and the spectral index evolution is shown in Figure \ref{fig_mjd_r_x} (in Panels III and IV), alongside our \seqsplit{quasi-simultaneous} X-ray coverage.

Our ATCA coverage investigated the evolution of the jet. As the source decayed at X-ray wavelengths at the end of the main outburst (around MJD 58250) it was not detected in the radio (with a 3$\sigma$ upper limit of $\lesssim$120 mJy/beam). Once \mx\,started re-brightening it increased by a factor of $\sim$18 in the X-ray count rate accompanied by a factor $\gtrsim$6 rise in the radio intensity (at both 5.5 and 9 GHz; see Figure \ref{fig_mjd_r_x}) in $\sim$2.9 days. The radio spectral index  $\alpha \! \sim \! 0.3$ indicated that the compact jet brightened rapidly once the source began re-brightening in the X-rays. The source continued to brighten further, transited to the SS, and the compact jet was quenched by a factor of $\gtrsim$16 in the 5.5 GHz band in $\lesssim$2.6 days. 

After the first re-brightening began to decay, the source transited back to a HS. We obtained a dense \seqsplit{quasi-simultaneous} X-ray and radio coverage of this decay starting around MJD 58269. The nearly daily radio coverage shows the transition of the radio spectral index from a steep ($\alpha\!\lesssim\!-1$) to an inverted radio spectrum ($\alpha \! \sim \! 0.5$) over $\sim$8 days as the jet switched back on and re-brightened. The source then decayed further in the X-rays and the brightness of the compact jet dropped. 

We obtained six more ATCA observations (after MJD 58290) as \mx\,evolved further. The source was detected during all but one of these observations (only upper limits were available for MJD 58335). One of these detections was obtained as the second re-brightening transited back to the HS (around MJD 58301) where $\alpha \! \sim \! -0.3$. The other four detections were obtained when \mx\, was in the HS during the third and fifth re-brightenings and confirm the presence of a compact jet ($\alpha \!  \gtrsim \!  0$). 

\subsection{Investigating the disk-jet coupling behaviour}
\label{sect_lr_lx}

\begin{figure}
\centering
\includegraphics[scale=0.46]{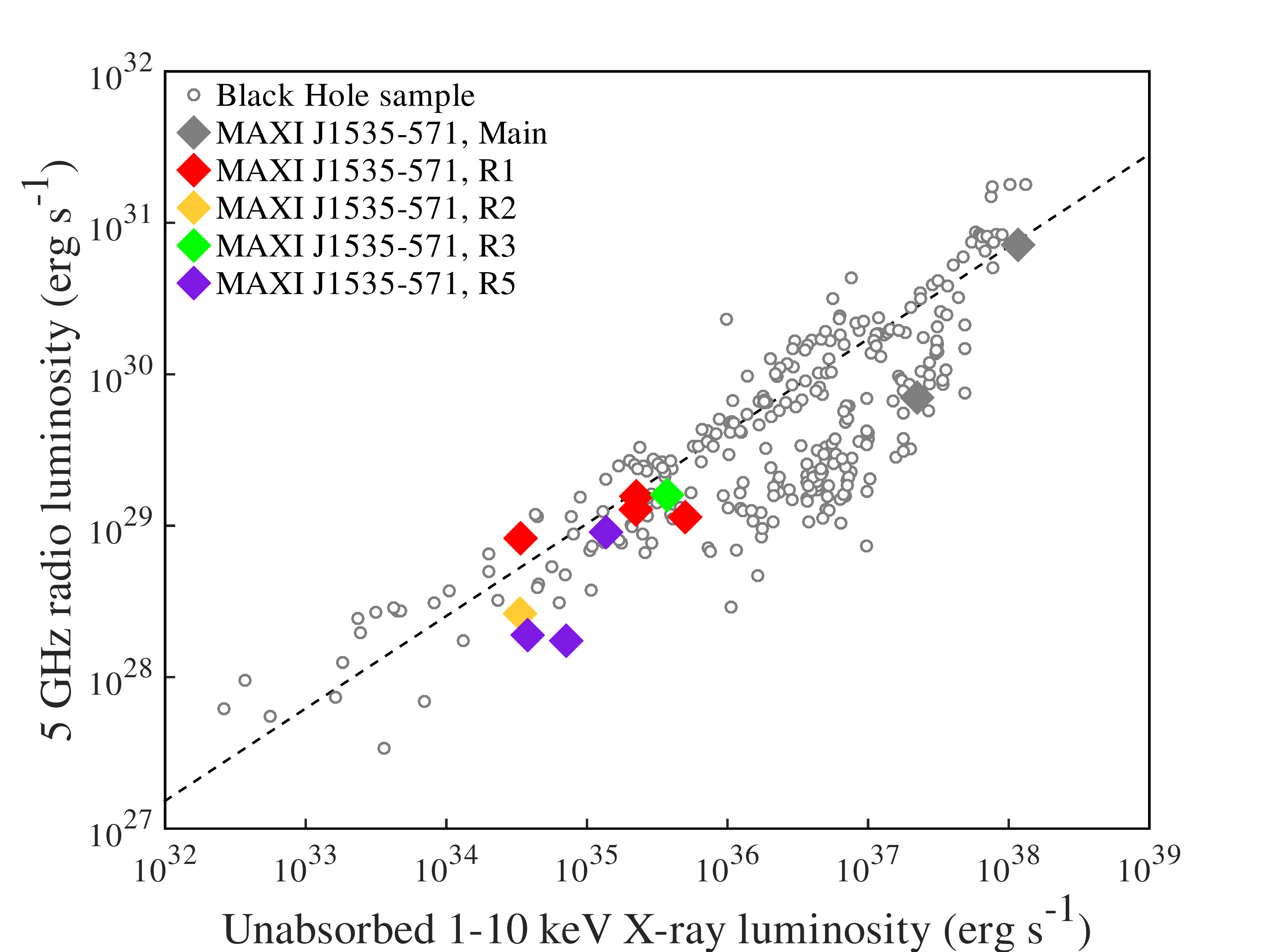}
\caption{The $L_\mathrm{R}$ and $L_\mathrm{X}$ data for a large sample of BH sources is shown \citep[using grey circles, adapted from][]{bahramian2018lrlx}. Data from \mx\,is shown for a distance estimate of 4.0  kpc \citep{chauhan2019maxi}, using diamonds, following the colour scheme used in Figure \ref{fig_lc_hid}. The dashed black line indicates the radio-loud track.}
\label{fig_lrlx}
\end{figure}

Using our quasi-simultaneous radio and X-ray coverage of \mx\,we have examined the $L_\mathrm{R}/L_\mathrm{X}$ relationship exhibited by the re-brightenings during their HS i.e., when a compact jet was active.

The HS X-ray spectra were fit using \textsc{XSpec} (v12.9.1m). The PC mode data were fit in the 0.5--10 keV energy range and the WT mode data in the 0.7--10 keV range.\footnote{The WT mode data are fit in the 0.7--10 keV range as the trailing charge in the spectra using the WT mode result in low energy spectral residuals below $0.7$ keV (see also http://www.swift.ac.uk/analysis/xrt/digest$\_$cal.php).} The equivalent hydrogen column density (\nh) was modelled using \textsc{WILM} abundances and \textsc{VERN} cross-sections \citep{wilms2000absorption,verner1996atomic}. Initially, all the HS spectra were modelled with an absorbed power-law, leaving the \nh\, free to vary for each fit. In order to determine the best-fit \nh\, we used the average \nh\, of these fits. This best fit \nh\, was found to be $(3.54\pm0.03)\times10^{22}$ cm$^{-2}$ and was fixed to this value for all further spectral fitting. The flux for the HS data, used for the $L_\mathrm{R}$ versus $L_\mathrm{X}$ investigation, was calculated using the the convolution model \texttt{cflux}. The luminosities reported here correspond to the 1--10 keV unabsorbed luminosities as this is the typical X-ray luminosity range for which the $L_\mathrm{R}$ versus $L_\mathrm{X}$ behaviour of sources is probed \citep[e.g.,][]{coriat2011radiatively,corbel2013universal,bahramian2018lrlx}\footnote{https://github.com/bersavosh/XRB-LrLx${\_}$pub}. Recent HI mapping of the region by \citet{chauhan2019maxi} suggested that the source is at a distance 4.0$\pm$0.2 kpc. Therefore, we have used a distance of 4.0 kpc to calculate the source luminosities.

The relationship between the $L_\mathrm{R}$ and $L_\mathrm{X}$ is probed at a radio frequency corresponding to 5 GHz. Therefore, we have extrapolated our observations to determine the flux expected at 5 GHz using the measured spectral index $\alpha$. Our X-ray and radio coverage of \mx\,is not strictly simultaneous. To ensure that we compare our $L_\mathrm{R}$ with simultaneous $L_\mathrm{X}$ data we have interpolated between the known X-ray luminosities (assuming the luminosity in logarithmic values) to obtain the required simultaneous $L_\mathrm{X}$.

The \mx\, data plotted in the $L_\mathrm{R}/L_\mathrm{X}$ plane is shown in Figure \ref{fig_lrlx}, using coloured diamonds, following the same colour scheme used in Figure \ref{fig_lc_hid}. For comparison with other BH sources we have adapted the data from \citet[][shown using grey circles]{bahramian2018lrlx}. Due to the lack of data we could not constrain the behaviour of \mx\, and the data could be consistent with both the radio-loud as well as radio-quiet tracks. However, the main outburst data appears to favour the radio-quiet track \citep[see][]{russell2019maxi}.

\section{Discussion}
\label{sect_disc}
\mx\, exhibited five re-brightenings after its main outburst ended and before it was Sun constrained. We carried out the first ever reported dense complementary X-ray and radio campaign to understand the relation between the accretion inflow and jet outflow in BH LMXBs during multiple rapidly evolving re-brightenings.


We found that although the first two re-brightenings transited  to the SS (at the same hardness traced out by the main outburst) they do not appear to exhibit the obvious hysteresis typically exhibited by most BH systems. This evolution of the re-brightenings is different from that exhibited by the main outburst which shows a clearly defined hysteresis loop. It is unknown why \mx\, does not exhibit the same hysteresis during its re-brightenings. This makes \mx\, different from GRS 1739$-$278 which is a BH LMXB that exhibited several clearly hysteretic re-brightenings \citep{yan2017detection}. \citet[][their Figure 7]{yan2017detection} have also shown the HID from two other sources (MAXI J1659$-$152 and XTE J1650$-$500) exhibiting re-brightenings which suggests that these sources also exhibited some hysteresis.

\citet{dunn2010global} compared the transition luminosities between the HS and SS in several BH LMXBs. Their study indicates that if the hard-to-soft state transition luminosity is relatively low it is likely that the reverse soft-to-hard transition will happen at a similar luminosity (see their Figure 10) which could result in little or no observed hysteresis. This is true for our source \mx. MAXI J1659$-$152 and XTE J1650$-$500, which also exhibited re-brightenings, do not have well constrained transition luminosities but they seem to transit at luminosities similar to \mx\, and, therefore, although they do exhibit some hysteresis there is not much difference in the transition luminosities in and out of the HS. However, GRS 1739$-$278 transited to the SS at a luminosity $\sim$10 times greater that observed for \mx\, and transited back to the HS at a luminosity similar to that observed for \mx, exhibiting relatively strong hysteretic behaviour. Thus, the re-brightenings in GRS 1739$-$278 evolve differently from those observed for \mx.

The third re-brightening of \mx\, was found to transit out of the HS but did not exhibit a hardness corresponding to that of the main outburst and first two re-brightenings in the SS. The fourth and partially observed fifth re-brightenings did not transit out of the HS. Studies of BH LMXBs show that the hard-to-soft transition luminosities can change for a given source and, therefore, this transition can take place at a range of luminosities \citep{belloni2006integral,dunn2008studying}. It is unknown what sets this transition luminosity. However, if this transition luminosity remained the same for all the re-brightenings in \mx\, it is no surprise that the source did not transit into the SS during its third, fourth, and fifth re-brightenings. This is because during these re-brightenings the source did not achieve the luminosity at which the first two re-brightenings were observed to reach the SS. We also note that the soft-to-hard transition luminosity observed for the main outburst and subsequent re-brightenings was not the same, as can be seen from the bottom panel of Figure \ref{fig_lc_hid}. Furthermore, the peak brightness of the re-brightenings was found to decrease with time. This may be indicative of an emptying reservoir of mass available for accretion onto the black hole.

We tracked the evolution of the jet during these re-brightenings. The source was accompanied by a compact jet during its HS as is typical for BH LMXBs. During the first re-brightening, the jet was quenched by a factor of $\gtrsim$16 in $\lesssim$2.6 days as \mx\, transited to the SS. In several BH LMXBs, radio flaring has been observed when the compact jet switches off during the hard-to-soft state transition. This flaring is thought to be caused by faster moving recent ejecta colliding with slower ejecta that is further away from the black hole \citep{fender2004towards,jamil2010ishocks}. No evidence of such flaring was found during our coverage (although we note that the hard-to-soft transition was only well monitored in the radio during the first re-brightening). This may suggest that the compact jet did not completely switch off or that the velocity at the transition to the SS was not sufficient to create internal shocks or no transient jet was launched.


We also obtained nearly daily ATCA coverage over a time scale of $\sim$8 days around MJD 58269 as MAXI J1535$-$571 transited back into the HS during the first re-brightening. We find that the radio spectral index $\alpha$ closely traced the observed evolution in the radio intensity and that the compact jet re-establishes as the source exhibits a soft-to-hard state transition. The compact jet in this source evolved from a steep to an inverted spectrum in $\sim$4 days, with the compact jet re-establishing over $\sim$8 days. This is very rapid compared GX 339$-$4, where the jet evolved from having a steep to an inverted spectrum in $>$10 days and with the compact jet taking $\sim$13 days to be re-established \citep{corbel2013formation}. This rapidly re-establishing compact jet supports the idea that the compact jet may not have switched off and instead the increase in the inflow (traced by the X-ray luminosity) causes the compact jet accelerating region to move closer to the BH and, therefore, the spectral break move to higher frequencies resulting in a brightening in our radio observations \citep[e.g.,][]{russell2013jet,russell2014accretion}.

We examined the evolution of the re-brightenings of \mx\,in the X-ray--radio luminosity plane. However, due to the lack of data from \mx\, the $L_\mathrm{R}/L_\mathrm{X}$ behaviour of the main outburst and subsequent re-brightenings could not be constrained to a specific $L_\mathrm{R}/L_\mathrm{X}$ track although the main outburst appears to slightly favour the radio-quiet track.

\section*{Acknowledgements}
AP and RW are supported by a NWO Top Grant, Module 1, awarded to RW. TDR is supported by a Netherlands Organisation for Scientific Research (NWO) Veni Fellowship. JCAM-J is the recipient of an Australian Research Council Future Fellowship (FT140101082), funded by the Australian government. GRS and AJT acknowledge support from an NSERC Discovery Grant (RGPIN-06569-2016). AJT acknowledges support from an Natural Sciences and Engineering Research Council of Canada (NSERC) Post-Graduate Doctoral Scholarship (PGSD2-490318-2016). The Australia Telescope Compact Array is part of the Australia Telescope, which is funded by the Commonwealth of Australia for operation as a National Facility managed by CSIRO.

 \newcommand{\noop}[1]{}

\end{document}